\documentclass[a4paper,11pt]{article}
\pdfoutput=1 

\usepackage{jinstpub} 

\usepackage{booktabs}
\usepackage{calc}
\usepackage[numbers]{natbib}

\usepackage{amsmath}
\usepackage{tabularx}
\usepackage[separate-uncertainty=true]{siunitx}
\usepackage{upgreek}
\DeclareSIUnit\eV{\ensuremath{\mathrm{eV}}}
\DeclareSIUnit\clight{\text{\ensuremath{c}}}

\usepackage{arydshln}
\usepackage{bigdelim}

\newcommand{\KrM}{\texorpdfstring{\ensuremath{^{83\mathrm{m}}\mathrm{Kr}}}{83mKr}}

\newcommand{\Rb}{\texorpdfstring{\ensuremath{^{83}\mathrm{Rb}}}{83Rb}}
\newcommand{\oldkryptonmode}{WGTS Loop Only Mode}
\newcommand{\newkryptonmode}{Double Injection Mode}

\newcommand{\tritiumcapillary}{tritium injection capillary}
\newcommand{\tritiumpluskryptoncapillary}{tritium+krypton injection capillary}

\newcommand{\oldindex}{WLO}
\newcommand{\newindex}{DI}

\title{Operation Modes of the KATRIN Experiment Tritium Loop System using $^\mathrm{83m}$Kr}

\newcommand{\etp}{Institute of Experimental Particle Physics~(ETP), Karlsruhe Institute of Technology~(KIT), Wolfgang-Gaede-Str. 1, 76131 Karlsruhe, Germany}
\newcommand{\tlk}{Tritium Laboratory Karlsruhe~(TLK),~Institute for Astroparticle Physics~(IAP), Karlsruhe Institute of Technology~(KIT), Hermann-von-Helmholtz-Platz 1, 76344 Eggenstein-Leopoldshafen, Germany}
\newcommand{\muenster}{Institut f\"{u}r Kernphysik, Westf\"alische Wilhelms-Universit\"{a}t M\"{u}nster, Wilhelm-Klemm-Str. 9, 48149 M\"{u}nster, Germany}
\newcommand{\npi}{Nuclear Physics Institute of the CAS, v. v. i., CZ-250 68 \v{R}e\v{z}, Czech Republic}
\newcommand{\washington}{Center for Experimental Nuclear Physics and Astrophysics, and Dept.~of Physics, University of Washington, Seattle, WA 98195, USA}
\affiliation[a]{\tlk}
\affiliation[b]{\washington}
\affiliation[c]{\muenster}
\affiliation[d]{\etp}
\affiliation[e]{\npi}

\author[a,b,1]{Alexander~Marsteller \note{Corresponding author}}
\author[c]{Matthias~B\"{o}ttcher}
\author[a]{Beate~Bornschein}
\author[b]{Sanshiro~Enomoto}
\author[d]{Caroline~Fengler}
\author[e]{Ond\v{r}ej~Lebeda}
\author[d]{Moritz~Machatschek}
\author[a]{Florian~Priester}
\author[e]{Jan~R\'{a}li\v{s}}
\author[a]{Marco~R\"{o}llig}
\author[a]{Carsten~R\"{o}ttele}
\author[a]{Magnus~Schl\"{o}sser}
\author[e]{Michal~\v{S}ef\v{c}\'{i}k}
\author[a]{Michael~Sturm}
\author[e]{Drahoslav~V\'{e}nos}

\emailAdd{amarst@uw.edu}

\abstract{
The KArlsruhe TRItium Neutrino (KATRIN) experiment aims to search for the effective electron antineutrino mass with a sensitivity of \texorpdfstring{\SI{0.2}{\eV}}{0.2 eV} (\texorpdfstring{\SI{90}{\percent}~}{\% }C.L.). 
In order to achieve this goal, KATRIN measurement phases focusing on the neutrino mass search are alternated with phases of investigations of systematic effects. 
During these phases, metastable \KrM{} is used as a calibration source.
The monoenergetic conversion electrons emitted accompanying the decay of \KrM{} allow a direct access to the starting conditions of \texorpdfstring{$\upbeta$}{beta}-electrons produced inside the windowless gaseous tritium source (WGTS) of KATRIN. 
To make use of \KrM{} in the WGTS, the Tritium Loop System, which provides a stable flow of tritium to the WGTS, needs to be operated in special modes.
This paper focuses on the technical implementation of these modes and their performance with regard to the achievable \KrM{}-rates, gas densities, and gas compositions inside the WGTS.
}


\begin{document}
\maketitle
\flushbottom

\section{Introduction}
The discovery of neutrino flavour oscillations around the turn of the last millennium~\cite{Fukuda1998,Ahmad2002} demonstrated that the neutrinos must be massive unlike originally assumed in the Standard Model of Particle Physics.
The differences of the squares of the three neutrino mass eigenvalues  were probed by neutrino-oscillation experiments with high precision.
However, the absolute scale of the neutrino mass remains one the of fundamental questions in (astro-) particle physics and cosmology.

A direct way of assessing the neutrino mass, which is independent of cosmological models or the mass nature of the neutrino, is the high-resolution spectroscopy or calorimetry of elements undergoing $\upbeta$-decay or electron capture \cite{Formaggio2021}.
The KArlsruhe TRItium Neutrino (KATRIN) experiment aims to measure the neutrino mass with a sensitivity of $\SI{0.2}{\eV\per\clight\squared}$ from the spectrum of tritium $\upbeta$-electrons \cite{Aker2021}. 
The experiment combines a high resolution ($\mathcal{O}(\si{\eV})$) Magnetic Adiabatic Collimation with Electrostatic (MAC-E)-Filter type spectrometer and an ultra-high luminosity molecular tritium source.

After the successful commissioning of the experiment with tritium in 2018 \cite{Aker2020a}, the KATRIN collaboration published the first upper limit for the neutrino mass of \SI{1.1}{\eV\per\clight\squared} in the following year from a data set of reduced source activity and a measuring period of only four weeks \cite{Aker:2019uuj}.
Recently, the first sub-\si{\eV} result was published and the upper limit was further decreased to \SI{0.8}{\eV\per\clight\squared} \cite{KNM2}. 
As the total foreseen measurement time of \SI{1000}{days} of beta-spectroscopy progresses, the statistics improves. 
At the same time the understanding and the quantification of systematic effects is needed to achieve the challenging final sensitivity. 

One key systematic effect is related to the starting potential of the $\upbeta$-electrons in the source. 
This potential is provided by a cold and strongly magnetized plasma in the tritium source. 
The electron/ion generation is linked to the tritium gas profile in the tritium source. 
This gas profile is experimentally characterized by the integral of its molecule density along the beam, the so called column density $\rho d$.
Strong magnetic fields of $B_{\mathrm{src}}=\SI{2.5}{\tesla}$ in the source confine the electron movement into the longitudinal direction. 
Ions behave similarly, but have a comparably larger radial drift probability due to their larger mass. 
The boundary conditions of this plasma are defined by the grounded beam tube and a source potential terminating gold plate.
Spatial inhomogeneities and temporal fluctuations in the plasma directly affect the starting potential of the $\upbeta$-electrons and thus the response function of the experiment, which directly affects the uncertainty on the neutrino mass.

The most effective way to determine the impact on the experiment's response function is the co-circulation of traces of metastable \KrM{} \cite{Venos2018} within the tritium gas.  
The spectrum of \KrM{} consists of discrete mono-energetic lines of the internal conversion and Auger electrons with natural linewidths and thermal Doppler-broadening.
Therefore, any shift or broadening of these lines reflects the spatial and temporal inhomogeneities in the plasma potential.
Calibration measurements using \KrM{} have been performed in the past during the first two KATRIN neutrino mass campaigns \cite{Aker:2019uuj,KNM2}. 
However, the optimal source conditions for the \KrM{} calibration ($T=\SI{100}{\kelvin}$,  $\rho d= \SI{1.475e17}{\per\centi\meter\squared}$ \cite{Machatschek2021}) and for the actual tritium neutrino mass scans of $\upbeta$-electrons ($T=\SI{30}{\kelvin}$, $\rho d= \SI{1.11e17}{\per\centi\meter\squared}$ \cite{Aker:2019uuj} / \SI{4.23e17}{\per\centi\meter\squared} \cite{KNM2}) are different.
For operation with tritium, the source is ideally operated at an as low as possible temperature of \SI{30}{\kelvin}, at which tritium condensation doesn't yet occur.
At this lowest possible operation temperature of \SI{30}{\kelvin}, \KrM{} would freeze out, thus requiring the operation of the source at higher temperatures above \SI{80}{\kelvin}. 
In the initial measurements with \KrM{}, the source was operated at \SI{100}{\kelvin}.
Observed plasma effects in the calibration may therefore not represent the actual source potential scenario under measurement conditions, which leads to additional uncertainties. 
For this reason, a new source temperature of $T=\SI{80}{\kelvin}$ and a novel loop-operation concept have recently been defined.
This represents the optimal compromise at which a column density of $\rho d=  \SI{3.75e17}{\per\centi\meter\squared}$ can be established for both the calibration and beta-scan modes.

In this paper, we introduce the working principle and the technical implementation of the ultra-luminous windowless gaseous tritium source, its tritium circulation loop and the interface to the \KrM{} generator based on \Rb{}.
We will describe the operational modes of the tritium source using \KrM{} and how the calibration measurements can be performed under the same conditions as during the neutrino mass measurements. 
Finally, we demonstrate the performance of the system and its impact on the quality of high-resolution spectroscopic measurements of monoenergetic \KrM{} lines.

\section{Experiment Setup}
The KATRIN experiment \cite{Aker2021} is located at the Karlsruhe Institute of Technology (KIT). 
The decision for this location was to a large part since the Tritium Laboratory Karlsruhe (TLK) is also situated there. 
As a unique facility it can provide the necessarily large amounts of high-purity tritium \cite{Doerr2005} required for the operation of the KATRIN experiment.

For the sake of brevity, in this text the term "krypton" is used to refer to a mixture of 
the metastable \KrM{} and {$^{83}$Kr}, and not $^{\mathrm{nat}}$Kr of naturally occurring isotope composition, which is not used in our experiment. 

\subsection{Windowless Gaseous Tritium Source}
\begin{figure}
    \centering
    \includegraphics{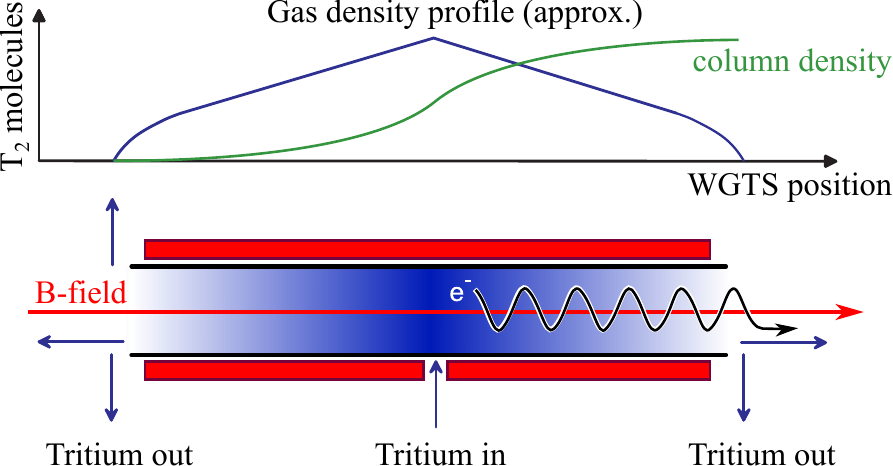}
    \caption{Windowless Gaseous Tritium Source (WGTS) schematic. 
    Gas density profile and column density (integral of gas density along the B field axis) are not to scale.}
    \label{fig:wgts_schematic_gas_flow}
\end{figure}

The KATRIN experiment uses a so-called Windowless Gaseous Tritium Source (WGTS) \cite{Aker2021}. 
This source essentially consists of a tritium gas column inside a \SI{10}{\meter} long tube of \SI{90}{\milli\meter} inner diameter, which is coupled to a two phase cooling system \cite{Grohmann2008}.
This allows for source operation at temperature levels of \SI{30}{\kelvin}, \SI{80}{\kelvin}, or \SI{100}{\kelvin} (using either neon, nitrogen, or argon as coolant).
Gas is injected into the WGTS via circumferential orifices from an injection chamber surrounding the central part of the WGTS beam tube.
Two injection capillaries (\tritiumcapillary{} and \tritiumpluskryptoncapillary{}) are connected to the injection chamber.
Both are thermally coupled to the beam tube over a length of \SI{5}{\meter}.
The whole source tube is installed inside the bore of a superconducting magnet system.
Attached to either end of the source tube there are two differential pumping ducts  (DPS1-F1/2 and DPS1-R1/2), each equipped with four or two turbomolecular pumps. 
Continuous injection of gas in the middle of the beam tube and pumping off at both ends leads to the formation of a gas density profile inside the source tube.
The experimentally accessible parameter of this density profile is its integral along the length of the WGTS, the so called column density (\autoref{fig:wgts_schematic_gas_flow}). 
The main requirements on the WGTS column density are the 24/7 operation with tritium at a purity \SI{>95}{\percent} and a stability (pressure and temperature) \SI{<0.1}{\percent}. 
The temperature stability is covered via the WGTS cryogenic system \cite{Grohmann2009}, while pressure stability and composition are maintained by the Tritium Loop System \cite{Priester2015}. 
Appropriate operation of both systems provides sufficiently stable column density \cite{Aker2021}.

\subsection{Tritium Loop System}
\begin{figure}[t]
    \centering
    \includegraphics{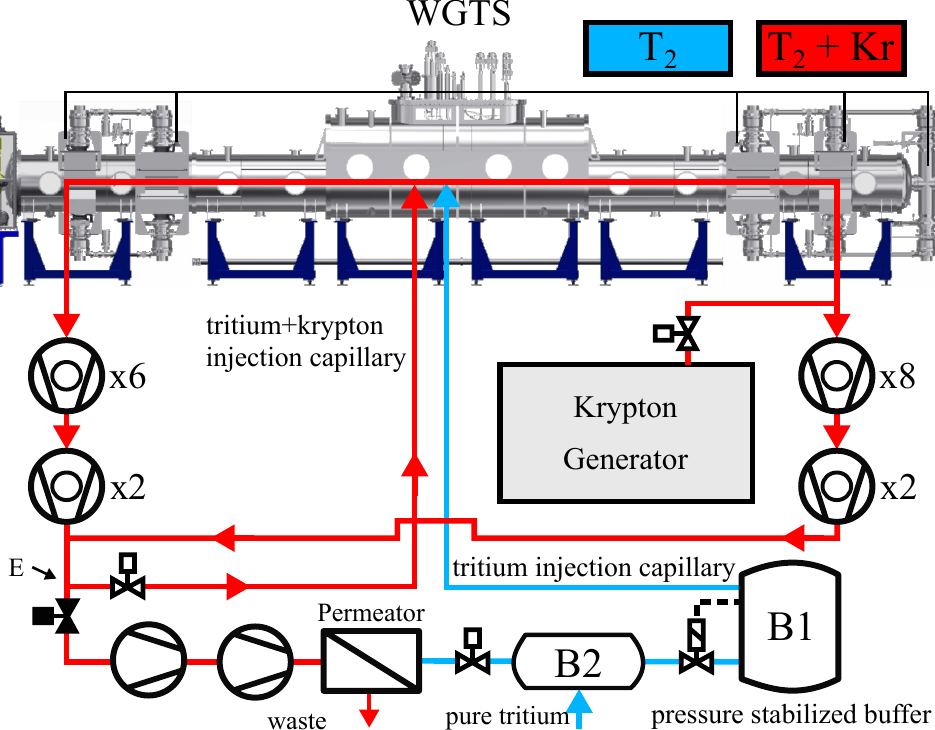}
    \caption{Schematic of the inner loop.}
    \label{fig:loop_schematic}
\end{figure}

The main task of the Tritium Loop System is the supply of KATRIN with tritium of high purity (\SI{>95}{\percent}) at a stable throughput of up to \SI{\approx 40}{\gram~tritium\per day}, as well as the reduction of tritium gas flow from the source to the KATRIN spectrometers by 14 orders of magnitude \cite{Aker2021,Marsteller2021}.
The flow reduction is achieved by the differential pumping sections DPS1 and DPS2 as well as by the cryogenic pumping section CPS \cite{Aker2021}.
The required high flow rate of tritium can only be achieved in a closed loop operation \cite{Priester2015,Sturm2021}. 
\autoref{fig:loop_schematic} shows a simplified flow scheme of the Inner Loop System. 

Tritium is injected from a pressure stabilized buffer vessel via the \tritiumcapillary{} and injection chamber into the WGTS beam tube.
The gas is pumped out at both ends of the tube by the turbomolecular pumps (TMP) of the DPS1. 
One end of the WGTS beam tube is connected to the spectrometers via the DPS2 and the CPS, while the other end is connected to the Rear Section (RS). The RS houses the Rear Wall (RW), a gold coated stainless steel disk of \SI{145}{\milli\meter} outer diameter installed perpendicularly to the magnetic flux tube \cite{Aker2021}. 
The RW is used to define the source potential, as well as for calibration and monitoring. 
The fore vacuum of the TMPs directly attached to the pumping ducts is provided by four 2$^{nd}$ stage TMPs, a scroll and a metal bellows pump in series. 

The last pump provides the pressure gradient needed to make the gas permeate through a palladium membrane filter ("permeator") into an intermediate buffer vessel B2. 
From there the gas passes through a Laser Raman sampling cell \cite{Aker2020} and a regulation valve back into the pressure stabilized buffer vessel B1, completing the closed loop.
The palladium membrane filter can only be passed by hydrogen isotopologues and therefore filters out all non-hydrogen impurities \cite{Bornschein2005, Goto1970}. 
A small fraction (\SI{\approx 1}{\percent}) of gas is continuously extracted in front of the permeator in order to avoid impurity accumulation and subsequent blocking of this filter. 
The amount of extracted gas is continuously replaced with high purity tritium from a system of buffer vessels which serves as the interface to the TLK infrastructure.
Depending on the operation mode the \tritiumpluskryptoncapillary{} can be either used in simultaneously with the \tritiumcapillary{} or exclusively, to directly inject the gas from the exhaust side of the 2$^{nd}$ stage TMPs into the WGTS beam tube.
These different operation modes are described in detail in  \autoref{sec:operation_modes}. 
An overview over the components of the Inner Loop System is given in \autoref{tab:loop_comp_table}.

\begin{table}[b]
    \centering
    \caption{Key components of the KATRIN Inner Loop}
    \begin{tabularx}{\textwidth}{lX}
    \toprule
         Component              & Function \\ \midrule
         AgPd membrane filter   & gas purification (only H/D/T can permeate) \\
         Turbomolecular pump    & 14 MAG W2800 TMP providing main vacuum inside WGTS beam tube, four HiPace300 providing fore pressure, backed by an all metal scroll/bellows pumping train. \\
         Vessel                 & two-staged buffer vessels for highly stable gas flow towards WGTS \\
         Flow meter             & monitoring of gas amount flowing into WGTS \\
         Laser Raman system     & custom build system for monitoring of gas purity/composition prior to injection into WGTS \\
         Beam tube gas analyser & MKS HPQ3 residual gas analyser used to monitor the gas composition when no B-fields are present\\
         \bottomrule
    \end{tabularx}
    \label{tab:loop_comp_table}
\end{table}

\subsection{Krypton Generator}
\begin{figure}[t]
    \centering
    \includegraphics[width=0.5\textwidth]{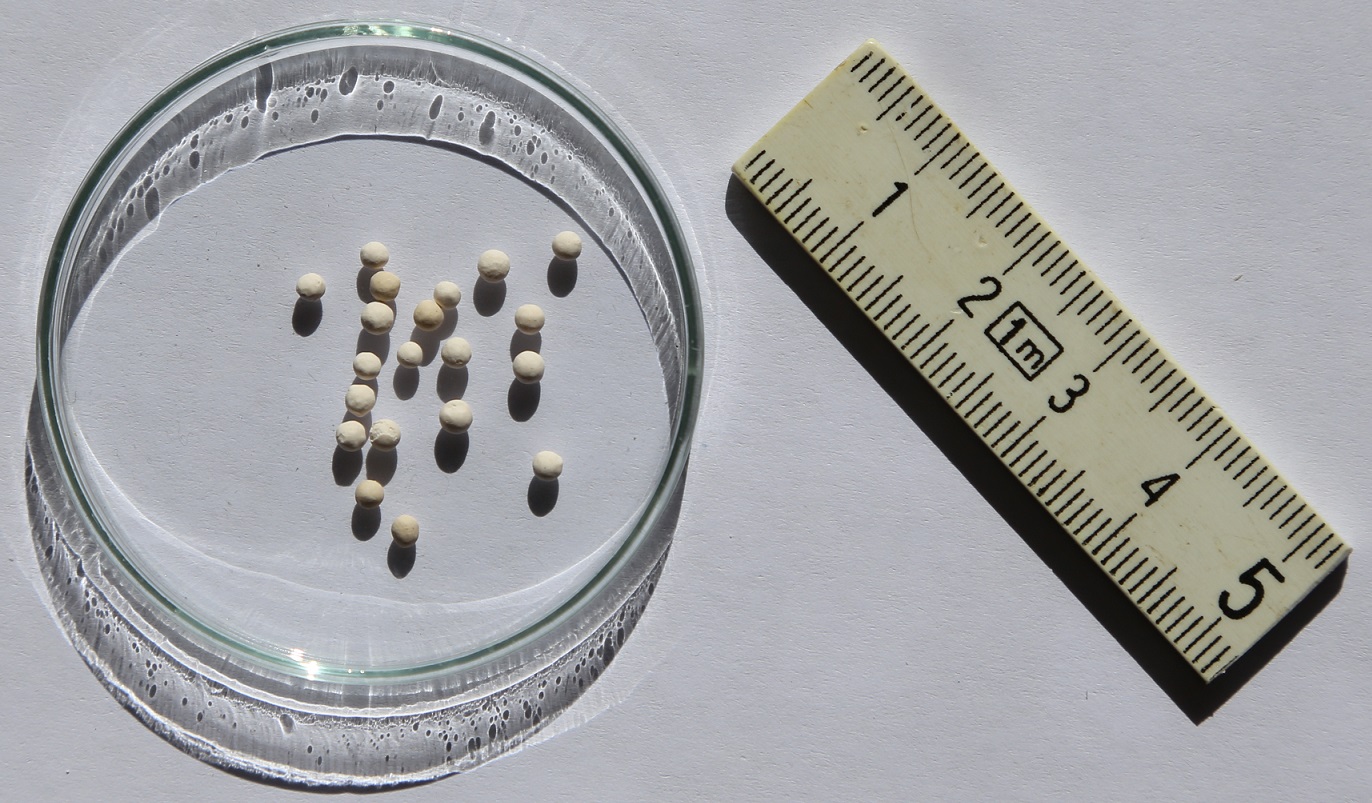}
    \caption{Picture of the \Rb{}/\KrM{} zeolite source, 20 spherules, total mass \SI{0.15}{\gram}.}
    \label{fig:Zeolite_Beads}
\end{figure}

The gaseous \KrM{} with a half-life of \SI{1.862\pm0.002}{\hour} \cite{Sentkerestiova2018} used by the KATRIN experiment is generated on-site via the decay of \Rb{} with a half-life of \SI{86.2\pm0.1}{\day} \cite{Audi2003}.
The \Rb{} has a probability of $f_{\mathrm{iso}} =\SI{74.5\pm3.8}{\percent}$ to decay into the \KrM{} isomeric state of $^{83}$Kr \cite{McCutchan2015, Venos2018}.
The parent \Rb{} is produced via $^{\mathrm{nat}}$Kr(p,xn) reactions using a pressurized gaseous target at the cyclotrons of NPI \v{R}e\v{z}. 
Due to the need of up to several \si{\giga\becquerel} of \Rb{}, the production rate of \Rb{} was gradually increased from \SIrange{14}{150}{\mega\becquerel} per irradiation hour \cite{Sefcik2020}.
The availability of monoenergetic electrons from the decay of gaseous \KrM{} depends in particular on the nature of the substrate used for the \Rb{} deposition. 
The substrate has to comply, among others, with two main requirements: 1) high emanation efficiency of \KrM{} and 2) as low as possible release of parent \Rb{} from the substrate. 
For KATRIN, we have decided to use natural cation exchange zeolite (molecular sieve - aluminosilicate material of type 5A by the company Merck). 
In \cite{Venos2005,Venos2014,Sentkerestiova2018} the deposition technique was described and the properties of the \Rb{}/\KrM{} radionuclide generator (further only "zeolite source") relevant for its use at the WGTS were determined. 
High and stable emanation efficiency of \KrM{} from the zeolite source (\SIrange{80}{90}{\percent} of the amount born in the \Rb{} decay), even in the presence of tritium, was demonstrated. 
The zeolite sources are prepared by depositing \Rb{} from aqueous solution into several tens of zeolite spherules, see \autoref{fig:Zeolite_Beads}. 
Experiences gained with the test version of the krypton generator setup are summarized in \cite{Sentkerestiova2018}. \\

\begin{figure}[t]
    \centering
    \includegraphics[width=0.7\textwidth]{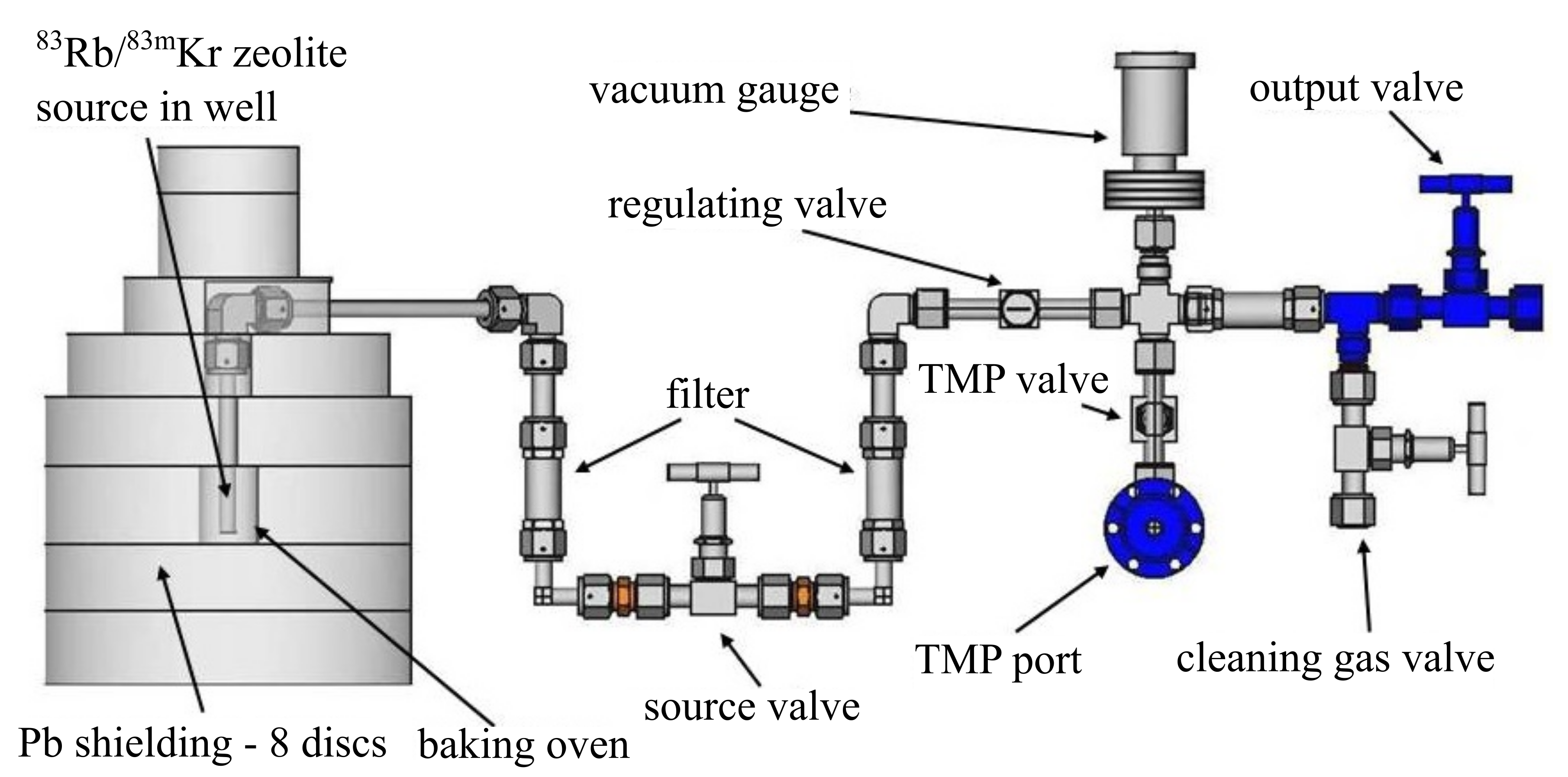}
    \caption{CAD rendering of the krypton generator setup.}
    \label{fig:Generator_Scheme}
\end{figure}

As KATRIN is an extremely sensitive, low background experiment, a lot of attention was paid to quantify the release of parent \Rb{} from zeolite.
The release was measured using zeolite sources containing \si{\giga\becquerel} levels of \Rb{}.
Strong ion bonds of rubidium in zeolite combined with two metal aerosol filters inserted into the generator setup resulted in a \Rb{} release equivalent to ca. \SI{20}{\milli\becquerel\per\day} observed at the generator output. 
The \Rb{}/\KrM{} generator setup for KATRIN depicted in \autoref{fig:Generator_Scheme} continuously supplies \KrM{} to the DPS1-F2 pump port, see \autoref{fig:loop_schematic}.
The generator's construction fulfils the requirements arising from the TLK licence for handling tritium in order to prevent tritium leakage. 
Therefore, only metal components with VCR\textsuperscript{\textregistered} or Conflat\textsuperscript{\textregistered} compatible, metal gaskets were used providing a single connection leak rate of \SI{<e-9}{\milli\bar\litre\per\second}.

A custom-made lead shielding around the zeolite source reduces the dose rate on the shielding surface to \SI{3}{\micro\sievert\per\hour} for a \Rb{} activity of \SI{2.5}{\giga\becquerel}. 
The whole generator setup with the shielding is located inside of a glove box in accordance with the safety measures maintained at the TLK.  

\vspace{0.5cm}
The generator setup consists of the main components, which are also shown in \autoref{fig:Generator_Scheme}: 
\begin{description}
    \item[Well] --- a welded stainless steel tube of \SI{12}{\cm} length which contains the zeolite spherules at its bottom.
    \item[Shielding] --- a \SI{160}{\kilo\gram} lead construction around the well with the zeolite source. It consists of four discs and four rings, allowing for the simple and rapid exchange of the well.
    \item [Ceramic oven] --- a custom-made oven located in the lowest ring of the shielding, enveloping the bottom of the well containing the zeolite spherules. 
    It enables baking of the zeolite up to \SI{200}{\celsius}. 
    The baking facilitates faster drying of zeolite in order to remove residual gases, in particular water vapor, from the zeolite prior to use of the generator.
    \item [Two sintered metal aerosol filters] --- these filters with pores of 
    \SI{0.5}{\micro\meter} diameter prevent potential aerosol and zeolite microparticles containing \Rb{} from spreading into the system.
    \item [Five valves] --- a source valve separates the zeolite source, a regulating valve allows for a rough control of the flow rate of the \KrM{} out of the generator setup, a valve connecting to a TMP allows for evacuation of the generator setup, a purge gas valve for the tube cleaning, and an output valve which connects the generator setup with the DPS1-F2 pumping port. 
    \item [Vacuum gauge] --- a vacuum gauge used to monitor the generator during pumpdown after installation, as well as during bakeout of the zeolite source.
\end{description}

\begin{figure}[t]
    \centering
    \includegraphics[scale=0.08]{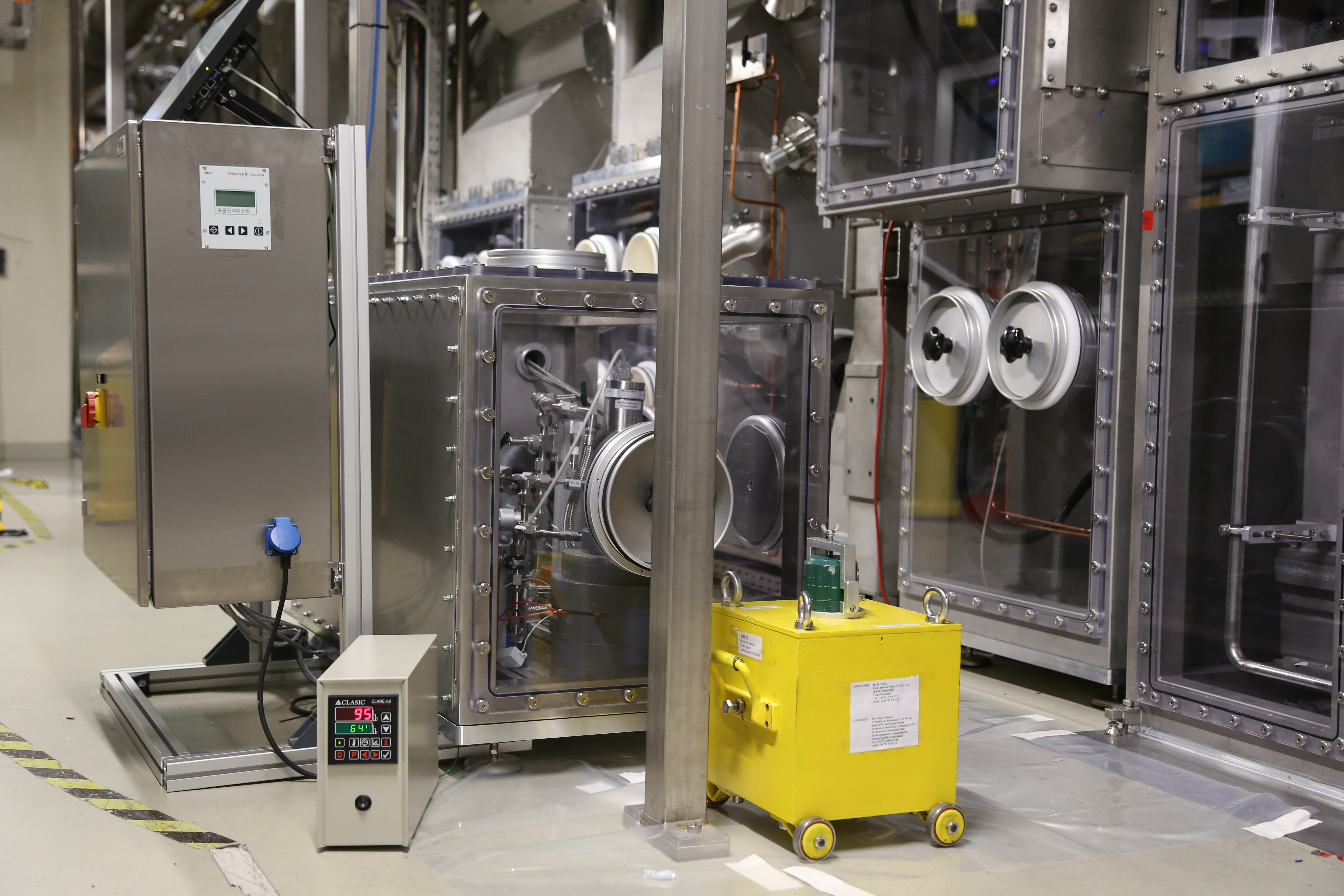}
    \caption{The \Rb{}/\KrM{} generator setup and yellow transport container. In
    the center, there is the 2$^{nd}$ containment box with the generator setup located inside. Electricity power distribution cabinet and controller of the
    baking oven are on the left.}
    \label{fig:Krypton_GeneratorTLK}
\end{figure}

The generator setup, connected to the WGTS in the TLK, is displayed in 
\autoref{fig:Krypton_GeneratorTLK}.
The krypton generator was successfully commissioned as a component of the KATRIN Gaseous Krypton Source (GKrS), providing the first results during the first transmission of electrons through the KATRIN beam line \cite{Arenz2018}. 
At that time the WGTS loop with the gas circulation system was not available yet. 
Instead, the \KrM{} from the generator was left to expand through DPS1-F2 into the  WGTS from where it streamed freely into the DPS2 and CPS beam tube and was pumped by the cold surface of the CPS.
Only the second commissioning step provided the standard loop system and admixture of D$_{2}$.

\section{Operation Modes}
\label{sec:operation_modes}

The successful operation of the KATRIN beamline with \KrM{} has been demonstrated in previous measurements \cite{Altenmueller2020,Arenz2018}.
However, these measurements were taken during the commissioning of the beamline, when the installation of the tritium loop system was not completed yet. 
This allowed for a setup with \KrM{} injection into the WGTS source tube without pumping by any of the TMPs, only cryogenically trapping the \KrM{} in the CPS.
After commissioning and operation of the WGTS with tritium, this configuration cannot be used anymore.

However, investigation of the systematic effects with impact of the neutrino mass determination requires an appropriate technical solution.
For this purpose, the tritium loop system of the KATRIN experiment is capable of two different operation modes incorporating \KrM{} at a source temperature of \SI{\geq 80}{\kelvin} to prevent it freezing-out. 
These modes are the so called \oldkryptonmode{} and \newkryptonmode{} \cite{Marsteller2020}.
In both modes, the \KrM{} is only present in traces; 
the number of \KrM{} atoms for an activity of \SI{10}{\giga\becquerel} is \SI{\approx 1e14}{atoms}, of which less than \SI{10}{\percent} is inside the source tube.
As the source tube contains \SI{\approx 1e19}{molecules} of T$_2$, the fraction of \KrM{} is less than \SI{1}{ppm} and does not have an impact on the gas dynamics of the circulating tritium gas.
This allows for a measurement of observables of the tritium plasma inside the source without significantly affecting the plasma itself.

\subsection{\oldkryptonmode{}}

In the \oldkryptonmode{}, the TMPs of the inner loop connected to the WGTS directly feed their exhaust gas back into the WGTS source tube via the \tritiumpluskryptoncapillary{}.
Initially, the pure tritium gas from buffer vessel B1, see \autoref{fig:loop_schematic}, is supplied into the inner loop.
Afterwards, the \KrM{} from the krypton generator is injected into the gas stream in front of one of these TMPs.
The tritium + \KrM{} mixture is then directly re-injected, which results in the accumulation of \KrM{} inside the circulating tritium gas.
In this manner, an amount of \KrM{} close to that of the activity of the \Rb{} parent is reached in the circulating gas.
The rate of \KrM{} electrons inside the WGTS in this mode is then given by that activity times the fraction of circulating gas inside the source tube. 
As such, this mode delivers the highest possible activity of \KrM{} in our loop configuration.

In addition to the direct reinjection of the circulating gas, it is necessary to supply a minuscule flow of tritium gas (\SI{\approx e-4}{\milli\bar\litre\per\second}) from the stabilized buffer vessel B1 via the \tritiumcapillary{} into this loop in order to maintain a stable gas column density inside the WGTS. 
This is due to the fact that only \SI{\approx99.9}{\percent} of gas is reinjected while the remainder is lost towards the DPS2 and RS, where it is pumped off  and sent to the TLK infrastructure for purification \cite{Marsteller2021}.

Furthermore, this operation mode suffers from impurity accumulation resulting from diffusion of air (N$_2$, O$_2$, Ar) through the polymer gaskets of inner loop TMPs as well as from outgassing (e.g. H$_2$, CH$_4$, CO$_2$, CO) of these pumps and the stainless steel walls and piping \cite{Priester2013}.
The generation of such outgassing impurities is enhanced by radiochemical processes induced by the $\upbeta$-decay of tritium, forming in particular a wide range of partially or fully tritiated methane species. 
In the neutrino mass measurement operation mode these impurities in the gas stream are filtered out by the permeator, see \autoref{fig:loop_schematic}.
As this permeator would filter out \KrM{} as well, it is not used in the \oldkryptonmode{}.
The accumulation of the impurities may affect the measurements as they slightly change the gas density inside the WGTS and may act as additional sources for electron scattering with an energy loss distribution different from tritium.

The process to change from the measurement configuration for neutrino mass measurements to the \oldkryptonmode{} requires among other steps a pump down of vessels B1 and B2, lasting \SI{>2}{\hour}.
The change from the \oldkryptonmode{} back to the neutrino mass measurement configuration necessitates a complete pump-down of the WGTS in order to remove the accumulated impurities, which requires \SI{\approx 3}{\hour}.

These disadvantages are avoided by implementation of a new operation mode, the \newkryptonmode{}.

\subsection{\newkryptonmode{}}
In contrast to the \oldkryptonmode{}, the \newkryptonmode{} is more complex.
Instead of a single gas loop, it consists of two intertwined loops.
One of these loops is the same as that of the \oldkryptonmode{}.
The other loop involves the permeator for the tritium purification.

Pure tritium gas is injected from the pressure stabilized buffer vessel, see \autoref{fig:loop_schematic}, via the \tritiumcapillary{} into the WGTS source tube.
From there it is pumped out by the inner loop TMPs where \KrM{} is mixed into the gas stream similarly to the \oldkryptonmode{}.
In the next step, in contrast to the \oldkryptonmode{}, the gas stream is split into roughly equal parts (see point E in \autoref{fig:loop_schematic}).
One part is directly re-injected into the WGTS source tube via the \tritiumpluskryptoncapillary{}, like in to the \oldkryptonmode{}.
The other fraction is pumped through the permeator and cleaned of impurities before returning into the pressure stabilized buffer vessel.

As a result, significantly less impurities accumulate, and the gas composition (disregarding \KrM{}) as well as the stability of the gas column density in this mode are equal for neutrino mass measurements and \KrM{} calibration measurements.
However, as a side effect of the gas purification, \KrM{} is also removed from the gas stream.
It results in a lower rate of \KrM{}, and thus lower conversion electron rate inside the WGTS source tube.
It does however allow for a quick (\SI{\approx 30}{\minute}) switching between neutrino mass measurements and \KrM{} calibration without interruption of the gas circulation by filtering out or injecting the \KrM{} (switching between krypton generator output valve closed and open) as well.

\subsection{Comparison of Modes}
Both operation modes have distinct advantages which make them complementary to each other:
The \oldkryptonmode{} provides the highest possible rate of electrons from \KrM{}-decay, allowing for spectroscopy of the \KrM{}-lines with unparalleled statistical precision. 
Meanwhile the \newkryptonmode{} is almost identical to the conditions of the tritium $\upbeta$-spectrum measurement.
It also allows rapid switching between modes, enabling measurements using \KrM{} on short timescales.
By combining information gained from measurements in both modes - high precision together with perfect comparability to the neutrino mass measurement - the potentials and plasma properties of the WGTS can be understood and their effect on the neutrino mass analysis taken into account.

\section{Performance}
The critical performance parameters for the KATRIN \KrM{} operation modes are the achievable rates of \KrM{} conversion electrons, the stability of these rates, the gas composition, and the column density. 
The interplay of these parameters lead to the capability to perform high precision spectroscopy on the \KrM{}-lines, which allows for the determination of systematic effects in the KATRIN setup such as plasma parameters of the source. 

\subsection{\KrM{} Rates}
The term \textit{rate} in the following text denotes the measured event rate at the KATRIN focal plane detector (FPD) \cite{Aker2021}, which is used to count electrons that have passed the MAC-E filter, in contrast to the entire \textit{activity} of \KrM{} or the \Rb{} parent. 
Furthermore, for easier understanding, only the rate at a particular retardation voltage setpoint of the MAC-E filter of \SI{30464.5}{\volt} will be presented here.
At this retardation voltage the entire rate corresponding to the $L_3$-32 electron line (as well as a small contribution from the $M_{1,2,3,4,5}$-32 and $N_{1,2,3}$-32 lines) of the \SI{32}{\kilo\eV} \KrM{} transition is measured, but none of the $L_3$-32 electron line which has scattered on gas molecules inside the source is.

The rate which is achievable in the \KrM{} operation modes at any given retardation voltage is determined by the following factors:

\begin{itemize}
    \item the activity $A_{\Rb{}}$ of the \Rb{} parent,
    
    \item the branching ratio of \Rb{} to the isomeric state \KrM{} $f_{\mathrm{iso}} =\SI{74.5\pm3.8}{\percent}$ \cite{McCutchan2015}, 
    
    \item the emanation efficiency $\varepsilon_{\mathrm{em}} = \SI{81.3\pm7.1}{\percent}$ of \KrM{} from the \Rb{} parent \cite{Venos2014},
   
    \item the fraction $f_{\mathrm{gas}}$ of gas inside the WGTS source tube to the entire gas amount containing \KrM{}.
    Values for this are typically around \SI{1.24\pm0.06}{\percent} at a column density of \SI{3.75e21}{\per\meter\squared} and decrease towards lower column densities,
    
    \item the fraction of $L_3$-32 electrons $f_{\mathrm{scat}}$ that have not scattered on the tritium molecules inside the source and experienced the subsequent energy loss, which is \SI{61.6\pm0.4}{\percent} for the above retardation voltage and a column density of \SI{3.75e21}{\per\meter\squared}, 
    
     \item the transmission efficiency $\varepsilon_{\mathrm{tr}} = \SI{18.1\pm0.1}{\percent}$ for emitted \KrM{} electrons to hit the FPD due to the limited acceptance angle of the magnetic electron guiding,
    
    \item the relative intensity of the $L_3$-32 line $f_{L_3-32} = \SI{37.8\pm0.5}{\percent}$ and the fraction $f_{\mathrm{spec}} = \SI{11.4\pm0.1}{\percent}$ of the \KrM{} spectrum above the $L_3$-32 line \cite{Venos2018}, as KATRIN uses a high pass integrating spectrometer, and
    
    \item the estimated detection efficiency $\varepsilon_{\mathrm{det}} = \SI{91\pm3}{\percent}$ of the used FPD wafer for the \KrM{} $L_3$-32 line with a \SI{6}{\kilo\eV} region of interest.
    
\end{itemize}

The activity $A_{\KrM{}}$ inside the WGTS source tube, once an equilibrium between \KrM{} and \Rb{} is reached, is given by: 

\begin{align}
A_{\KrM{}} &= A_{\Rb{}} \cdot f_{\mathrm{iso}} \cdot \varepsilon_{\mathrm{em}}   \cdot f_{\mathrm{gas}} \cdot f_{\mathrm{ex}}\,,\\
&= A_{\Rb{}, \mathrm{eff}} \cdot f_{\mathrm{gas}} \cdot f_{\mathrm{ex}}\,,
\end{align}
where $A_{\Rb{}, \mathrm{eff}} = A_{\Rb{}} \cdot \varepsilon_{\mathrm{em}} \cdot f_{\mathrm{iso}}$ is an abbreviation expressing the \Rb{} activity which leads to emanation of \KrM{} into the gaseous phase. 
The factor $f_{\mathrm{ex}}$ describes the reduction of \KrM{} in the system due to partial purification of the gas in the \newkryptonmode{}. 
In the \oldkryptonmode{} it is 1.
This factor is derived by comparing the \oldkryptonmode{} equilibrium with a simple model assumption for the amount of \KrM{} inside the gas in the \newkryptonmode{}. 
This model is described by the differential equations for the change in \KrM{} amount $N_{\KrM{}}$ in both modes ($N_{\KrM{}, \mathrm{\oldindex}}$ for the \oldkryptonmode{}, $N_{\KrM{}, \mathrm{\newindex}}$ for the \newkryptonmode{}):

\begin{align}
    \dot{N}_{\KrM{}, \mathrm{\oldindex}} &= - N_{\KrM{}, \mathrm{\oldindex}} \cdot \lambda_{\KrM{}} + A_{\Rb{}, \mathrm{eff}}\,,\\
    \dot{N}_{\KrM{}, \mathrm{\newindex}} &= - N_{\KrM{}, \mathrm{\newindex}} \cdot (\lambda_{\KrM{}} + \frac{q_{\mathrm{ex}}}{n_{\mathrm{tot}}}) + A_{\Rb{}, \mathrm{eff}}\,,
\end{align}

with the decay constant $\lambda_{\KrM{}}$ of \KrM{}, the total gas amount $n_{\mathrm{tot}}$, and the extracted flow rate $q_{\mathrm{ex}}$.
These equations simplify in equilibrium (${\dot{N}_{\KrM{}, \mathrm{\oldindex}} = \dot{N}_{\KrM{}, \mathrm{\newindex}} = 0}$) to:

\begin{align}
    N_{\KrM, \mathrm{\oldindex}} &= \frac{1}{\lambda_{\KrM{}}} \cdot A_{\Rb{}, \mathrm{eff}}\,,\\
    N_{\KrM, \mathrm{\newindex}} &= \frac{1}{\lambda_{\KrM{}} + \frac{q_{\mathrm{ex}}}{n_{\mathrm{tot}}}} \cdot A_{\Rb{}, \mathrm{eff}}\,. \label{eq:alt_model}
\end{align}

Due to the design of the setup, every \KrM{} atom leaving the generator first passes the point of extraction (see label E in \autoref{fig:loop_schematic}) before entering the WGTS source tube for the first time.
This means that only a fraction of the \KrM{} released by the generator which is given by the additional factor of $\frac{q_{\mathrm{tot}} - q_{\mathrm{ex}}}{q_{\mathrm{tot}}}$ contributes to the amount of \KrM{} in the source tube, leading to a total correction factor of:

\begin{align}
    f_{\mathrm{ex}} &= \frac{N_{\KrM, \mathrm{\newindex}}}{N_{\KrM, \mathrm{\oldindex}}} \cdot \frac{q_{\mathrm{tot}} - q_{\mathrm{ex}}}{q_{\mathrm{tot}}}\,,\\
    f_{\mathrm{ex}} &= \frac{\lambda_{\KrM{}}}{\frac{q_{\mathrm{ex}}}{n_{\mathrm{tot}}} + \lambda_{\KrM{}}} \cdot \frac{q_{\mathrm{tot}} - q_{\mathrm{ex}}}{q_{\mathrm{tot}}}\,,
\end{align}

with the total flow rate through the WGTS source tube $q_{\mathrm{tot}}$.
The \KrM{} activity inside the WGTS source tube derived in this way translates to the expected FPD rate $R_{\mathrm{FPD, exp}}$ as follows:

\begin{align}
    R_{\mathrm{FPD, exp}} = A_{\KrM{}} \cdot  \left(f_{\mathrm{scat}} \cdot f_{\mathrm{L}_3\mathrm{-32}} + f_{\mathrm{spec}}\right)\cdot \varepsilon_{\mathrm{tr}} \cdot \varepsilon_{\mathrm{det}}\,.
\end{align}

\begin{figure}
    \centering
    \includegraphics{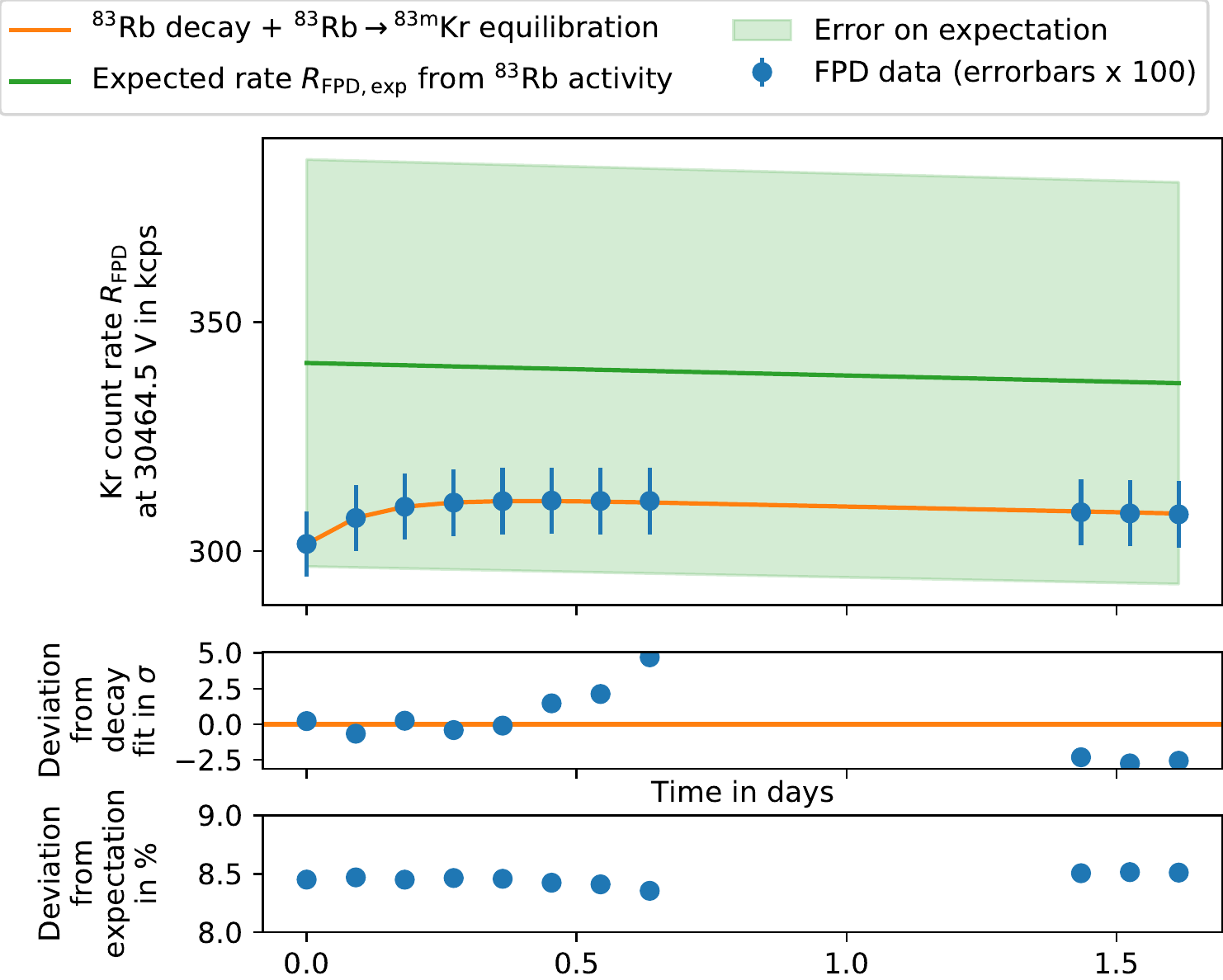}
    \caption{Rate stability in the \oldkryptonmode{}. Effects of \Rb{} decay and $\Rb{} \rightarrow {\KrM{}}$ production fitted to data with fixed decay constants taken from literature \cite{Venos2018,McCutchan2015,Audi2003,Sentkerestiova2018}.
    The uncertainty on the expectation is from all the uncertainties listed in the text. 
    The uncertainty on FPD rate data is statistical only.}
    \label{fig:wgts_only_mode_rate_stability_stacked}
\end{figure}

\begin{figure}
    \centering
    \includegraphics{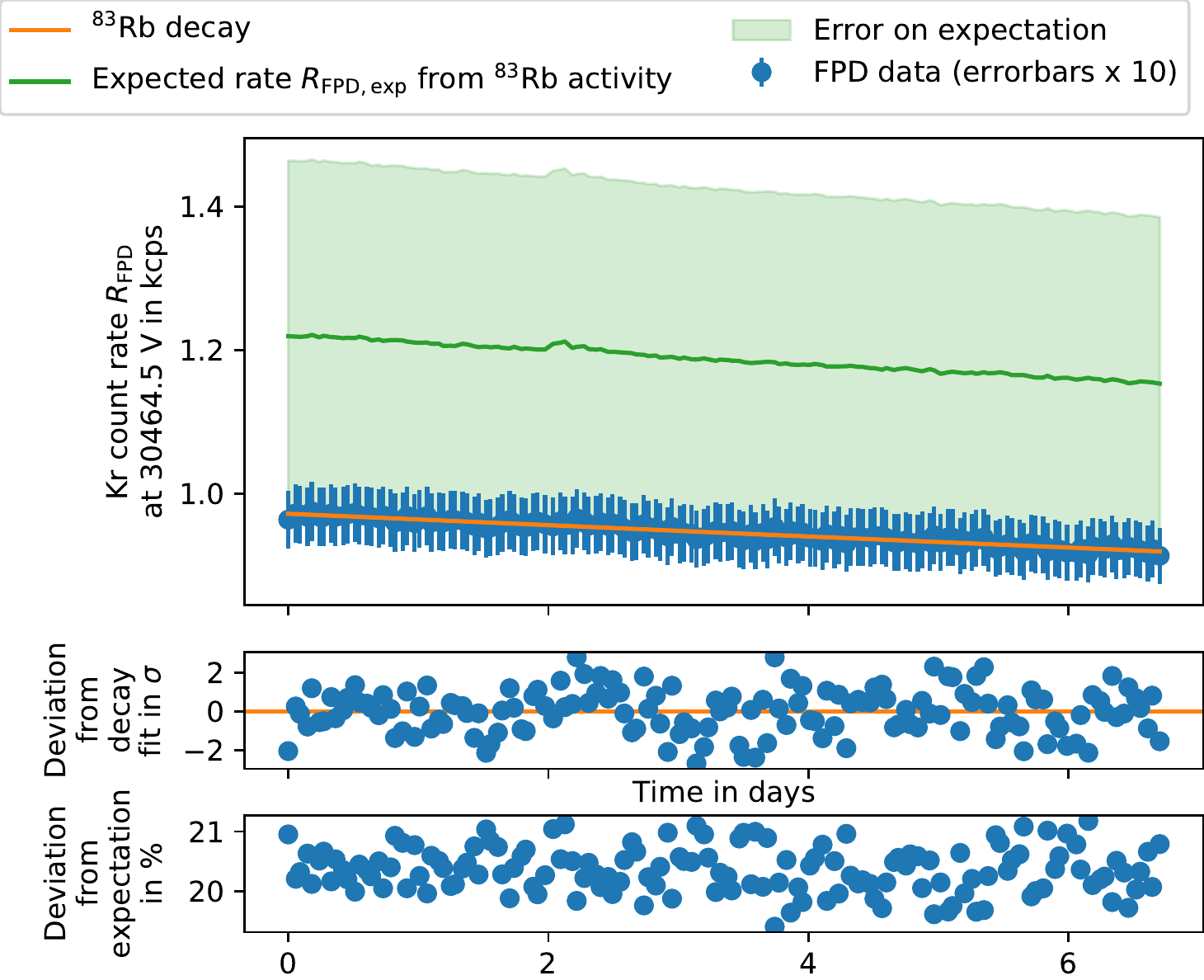}
    \caption{Expected versus measured FPD rate in the \newkryptonmode{}. }
    \label{fig:double_injection_mode_rate_stability_stacked2}
\end{figure}

Based on these considerations, the expected and measured \KrM{} rates for representative time periods of \oldkryptonmode{} and \newkryptonmode{} operation are shown in \autoref{fig:wgts_only_mode_rate_stability_stacked} and \autoref{fig:double_injection_mode_rate_stability_stacked2}, respectively.
The figures also contain exponential fits with fixed decay constants and free amplitudes, describing the rate evolution due to \Rb{} and \KrM{} decays, which are elaborated on further below.
For both operation modes, the expected and measured FPD rates agree well within their uncertainties.
The systematic shift in both modes is likely the result of the large systematic uncertainty on the gas dynamical properties such as the volumes needed to calculate $n_{\mathrm{tot}}$, or the emanation efficiency. 

Comparing the rates in both modes also immediately shows the drawback of the \newkryptonmode{}.
In exchange for the inclusion of the gas purification system, the rate is reduced by $f_{\mathrm{ex}} = \SI{1.86\pm0.28e-3}{}$. Therefore, at the same activity $A_{\Rb{}}$ of the \Rb{} parent, much longer measurement times are required in the \newkryptonmode{} compared to the \oldkryptonmode{}.

\begin{figure}
    \centering
    \includegraphics{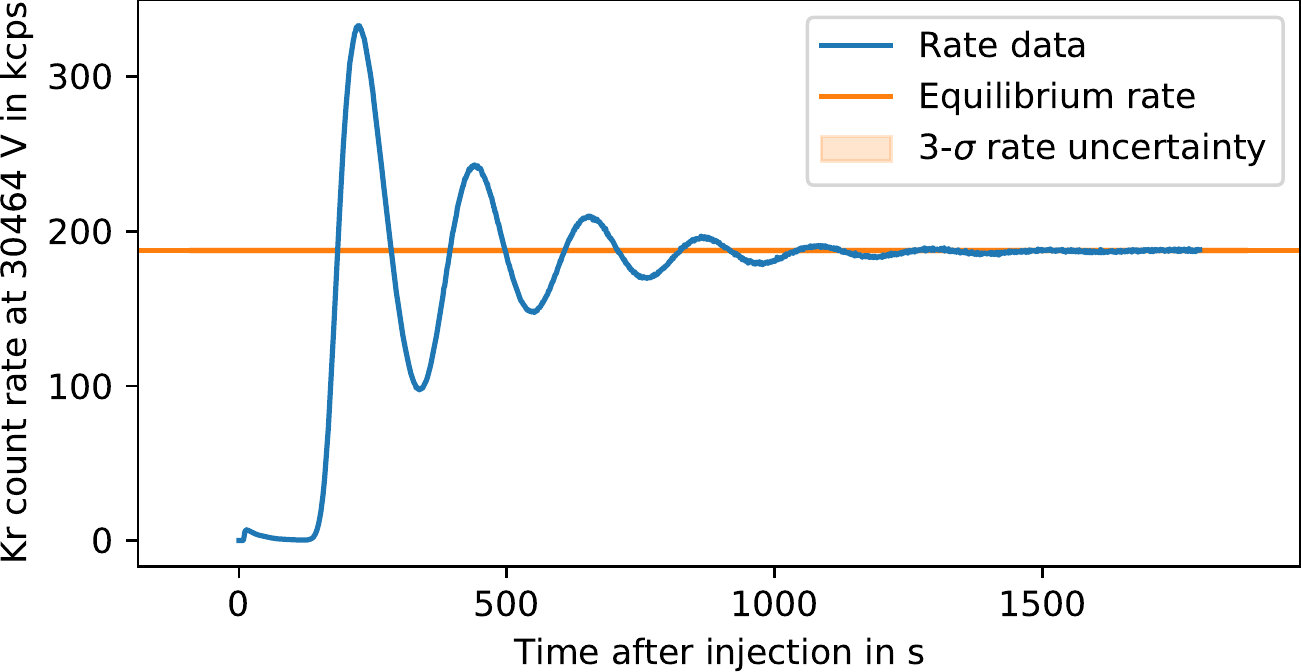}
    \caption{Time until rate stabilization in the \oldkryptonmode{}{}. The \SI{3}{\sigma} uncertainty band mostly overlaps with the equilibrium rate line.}
    \label{fig:tritium_plus_krypton_mode_krypton_injection}
\end{figure}

\autoref{fig:wgts_only_mode_rate_stability_stacked} also illustrates the slow attaining of the transient equilibrium between the \KrM{} and \Rb{} parent that takes several \KrM{} half-lives.
This effect can be suppressed, if \KrM{} is accumulated inside the krypton generator before starting the injection into the circulating tritium gas.
This is shown in \autoref{fig:tritium_plus_krypton_mode_krypton_injection}, where the FPD rate during an injection of accumulated \KrM{} into the circulating gas stream is displayed. 
The injected bunch of \KrM{} is circulated through the closed WGTS loop, leading to the oscillatory behavior, until diffusing and mixing of \KrM{} into the surrounding tritium result in achieving equal distribution in the entire loop within \SI{\approx 30}{\minute}.
From that moment onward, the stability of the measured rate is dominated by the decay of the \Rb{} parent of \SI{0.8}{\percent\per\day}, or the equilibration between the \KrM{} and the \Rb{} parent if the \KrM{} was not allowed to accumulate.

\begin{figure}
    \centering
    \includegraphics{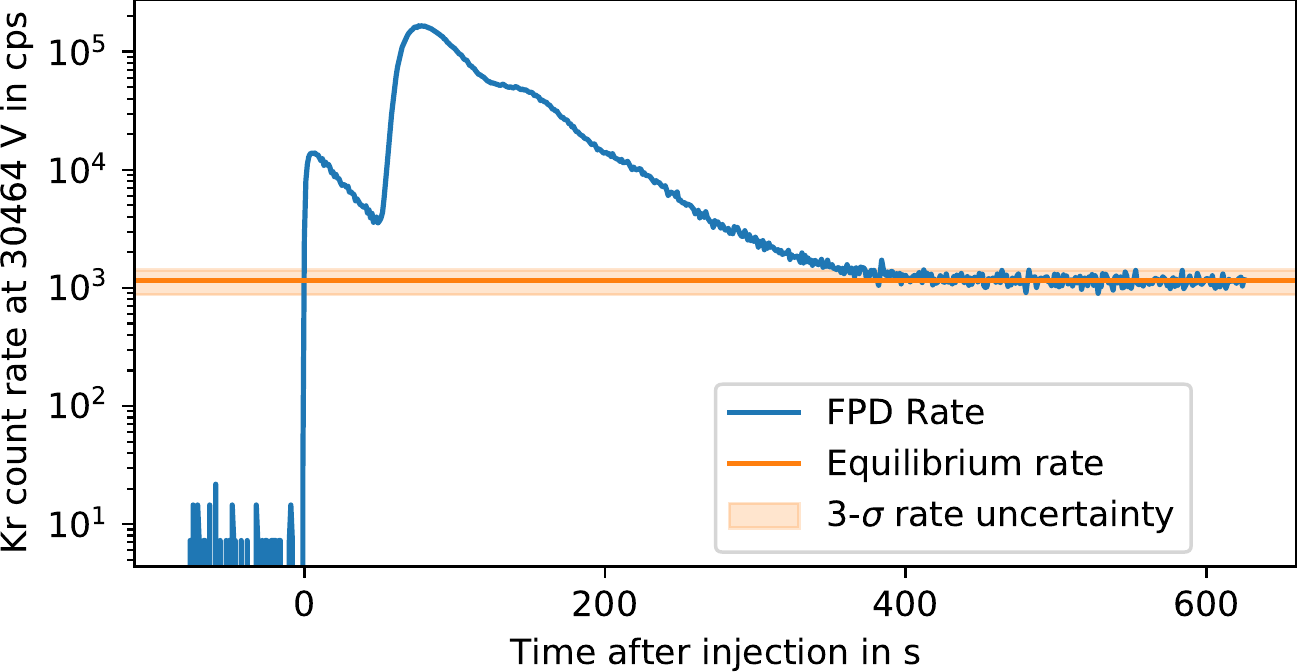}
    \caption{Time until rate stabilization in the \newkryptonmode{} }
    \label{fig:alternative_tritium_plus_krypton_mode_stabilization_time}
\end{figure}

In the \newkryptonmode{}, the behavior upon injection of \KrM{} is vastly different, as is depicted in \autoref{fig:alternative_tritium_plus_krypton_mode_stabilization_time}.
As the term related to gas exchange in \autoref{eq:alt_model} is much bigger than the decay constant $\lambda_{\KrM{}}$ of \KrM{}, no equilibrium between the \KrM{} inside the loop and the \Rb{} parent is allowed to develop.
Instead, an equilibrium between the production rate of \KrM{} by the \Rb{} parent and the filtration rate of \KrM{} by the gas purification system is established.
This oscillatory behavior after an initial injection is heavily dampened, and the equilibrium value is reached within \SI{< 10}{\minute}; much faster than in the \oldkryptonmode{}.

In both modes, the rate stability, disregarding the smooth decrease due to decay of the \Rb{} parent, is excellent.
By fitting two exponential functions with fixed decay constants for the \Rb{} and \KrM{} and free amplitudes to the data, the remaining fluctuations in the residuals were investigated.
For the \oldkryptonmode{}, the relative standard deviation of these residuals is \SI{0.050}{\percent} with an expectation of \SI{0.023}{\percent} from purely statistical rate fluctuations ($\chi^2_\nu / \text{d.o.f} = 5.46$ for fit with only statistical uncertainties).
Under the assumption of Gaussian fluctuations, we can derive from these values that the stability of our circulation system, disregarding the decay, is on the order of \SI{0.044}{\percent} in the \oldkryptonmode{}.
In comparison, the lower rates in the \newkryptonmode{} yield a much larger standard deviation of \SI{0.49}{\percent} with an expectation of \SI{0.42}{\percent} from purely statistical rate fluctuations ($\chi^2_\nu / \text{d.o.f} = 1.36$ for fit with only statistical uncertainties).
Deriving the stability for the \newkryptonmode{} as done above, we obtain a value on the order of \SI{0.25}{\percent} for the rate stability.

From June to August 2021, a \Rb{} source with an extraordinarily high activity of \SI{9.8}{\giga\becquerel} was used to enable high sensitivity spectroscopy of the $N_{2,3}$-32-lines of \KrM{}. 
Such a high activity source was necessary in order to achieve rates of \SI{\approx 30}{cps} for the $N_{2}$-32 line in the \newkryptonmode{}, i.e. of \SI{\approx 30}{\kilo cps} in the \oldkryptonmode{}.
The results of these spectroscopic measurements will be presented in an upcoming publication.

\subsection{Gas Composition}

Both krypton modes require a recirculation - partial in the case of the \newkryptonmode{} - of the gas pumped out of the WGTS source tube via TMPs. 
This has the disadvantage that, due to the internal construction of the TMPs, the gas stream on the high pressure side of the TMP is contaminated by gases other than hydrogen isotopologues or krypton.

The TMPs are not fully metal and are sealed with polymer gaskets. 
Therefore in addition to the outgassing of and exchange reactions inside the TMPs \cite{Priester2013}, a leakage flow with a rate on the order of \SI{<e-6}{\milli\bar\litre\per\second} per pump enters the system.
These impurities are undesired for operation of the krypton modes as they can slightly change the column density, plasma parameters, and serve as additional molecules for $\upbeta$-electrons to scatter on with different energy loss distributions compared to molecular tritium.

In order to investigate these impurities in-situ, a {MKS HPQ3} residual gas analyzer (RGA) is installed at the WGTS.
We refrain in the following from making quantitative statements as the RGA is operated in the stray magnetic field of the WGTS magnets, which significantly impacts the peak heights for different masses in a non-trivial fashion, and due to the inability to calibrate the RGA in-situ. 

\begin{figure}
    \centering
    \includegraphics{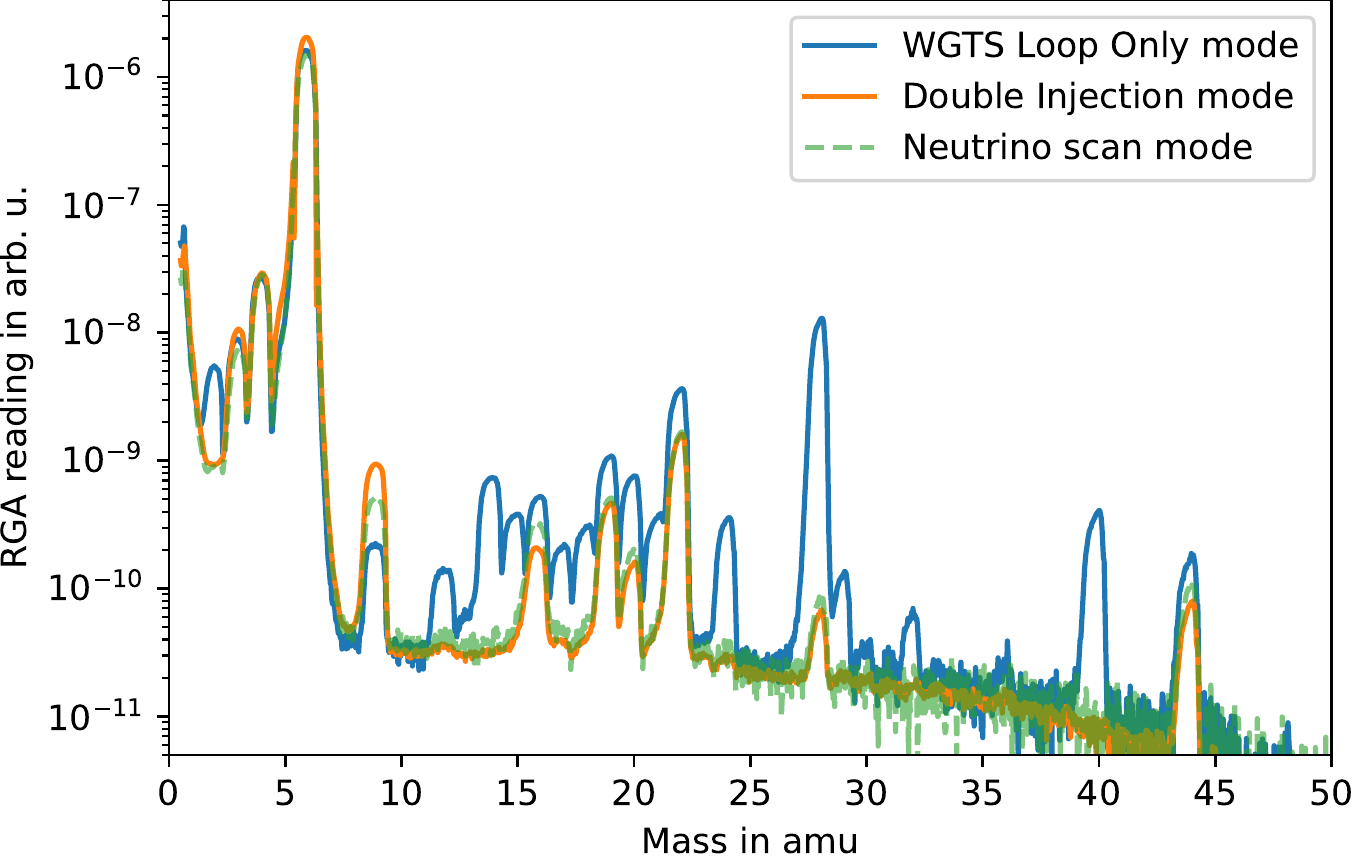}
    \caption{\textbf{Gas composition during circulation in \oldkryptonmode{} and \newkryptonmode{}.}
    All data sets were taken at similar total pressure and magnetic field settings. 
    In addition to the krypton operation modes, a spectrum for the normal neutrino scan mode without recirculation of gas is shown for comparison.
    All data are averaged over \SI{\approx 100}{} individual mass spectra to make the low intensity peaks visible in the RGA statistical noise.}
    \label{fig:rga_spectrum}
\end{figure}

Mass spectra for both the \newkryptonmode{} and the \oldkryptonmode{} measured with this RGA are shown in \autoref{fig:rga_spectrum} together with a spectrum of the operation mode without gas recirculation for comparison.
A list identifying the peaks in the mass spectra with potential gas species is given in \autoref{tab:rga_spectra_alt}.
Relative factors for the suppression of the peaks from \oldkryptonmode{} to \newkryptonmode{} are given as well.
However, these factors are only an estimation as most peaks are already close to the detection limit in the \oldkryptonmode{}, and thus only a lower limit can be set for the disappearing peak. 
This is further complicated by some peak shapes extending across several amu, and some fragmentation patterns overlapping other peaks such as with CO$_2$ and N$_2$ at mass 28.

The peaks visible in the \newkryptonmode{} are almost identical to those without gas recirculation, showing it to contain no significant amount of additional impurities.
The peaks present besides the hydrogen isotopologues are tritiated water, and CO as well CO$_2$ which are produced by the RGA or originate from the RS. 

Compared to the spectrum of the \newkryptonmode{}, the spectrum of the \oldkryptonmode{} contains several additional peaks.
These peaks are to a large extent due to air (nitrogen, oxygen, argon). 
However, the spectrum contains a comb-structure, in addition to the water related peaks, between mass 12 to 24, which can be identified with tritiated methane species and their fragments.
In general, these non-hydrogen isotopologue peaks are much more pronounced in the \oldkryptonmode{}, which is expected from a precursor experiment \cite{Priester2013} with a similar gas circulation configuration.

This is a natural consequence of the complete lack of gas purification in the \oldkryptonmode{}, as the peaks increase in time as the impurities accumulate.
In the \newkryptonmode{} however, an equilibrium between the impurity inflow rate and the gas extraction rate settles in at a level which is not visible with the RGA in our setup, showing a spectrum which is nearly identical to the mode without gas recirculation.

\begin{table}[ht]
    \centering
    \caption{
    \textbf{Observed peaks in the RGA measurements and possible corresponding gas species.}
    Listed are all peaks present in the \oldkryptonmode{} and it is noted which of these peaks are also visible in the \newkryptonmode{}.
    Details on the relative suppression factors are given in the text.}
    \begin{tabularx}{\textwidth}{rcXXXr}
    \toprule
         Mass in \si{\amu}  &  Possible gas species                 & Relative suppression & Visible in \newkryptonmode{} & & \\ \midrule
         1                  & H                                     &  - & Yes & \rdelim\}{7}{*}[\parbox{8cm-\tabcolsep-\widthof{$\Bigg]$}}{hydrogen and\\ helium \\ isotopologues}]\\ 
         2                  & H$_2$, D                              &  \num{4.3e+01} &Yes && \\
         3                  & H$_3$, HD, T, $^{3}$He                &  \num{8.1e-01} &Yes & & \\
         4                  & D$_2$, D$_2$, $^{4}$He                &  \num{9.2e-01} &Yes && \\
         5                  & DT                                    &  \num{5.1e-01} &Yes && \\
         6                  & T$_2$                                 &  \num{7.8e-01} &Yes && \\
         9                  & T$_3$                                 &  \num{2.0e-01} &Yes && \\ \cdashline{1-4} \noalign{\vskip 0.5ex}
         12                 & C                                     &  \num{>2.4e+01} &No & \rdelim\}{11}{*}[\parbox{8cm-\tabcolsep-\widthof{$\Bigg]$}}{tritiated water\\ and hydrocarbons}]& \\
         14                 & CH$_2$, CD, $^{14}$N                  &  \num{>2.0e+02} &No && \\
         15                 & CT, CH$_3$, CHD, $^{15}$N             &  \num{2.4e+01} &No && \\
         16                 & CH$_4$, CTH, CD$_2$, O                &  \num{3.0e+00} &Yes & & \\
         17                 & CH$_2$T, CH$_3$D, OH                 &  \num{3.2e+01} &Yes & & \\
         18                 & CH$_3$T, CT$_2$, H$_2$O, OD           &  \num{2.7e+01} &Yes && \\
         19                 & CHT$_2$, F, OT, HDO                   &  \num{2.5e+00} &Yes && \\
         20                 & CH$_2$T$_2$, Ar$^{++}$, HTO, D$_2$O   &  \num{5.7e+00} &Yes && \\
         21                 & CD$_2$T$_2$, DTO                      &  \num{1.5e+01} &No && \\
         22                 & CHT$_3$, T$_2$O                       &  \num{2.3e+00} &Yes && \\
         24                 & CT$_4$,                               &  \num{5.3e+01} &No && \\ \cdashline{1-4} \noalign{\vskip 0.5ex}
         28                 & CO, $^{14}$N$_2$                      &  \num{2.9e+02} &No, Fragment of CO$_2$ & \rdelim\}{7}{*}[\parbox{8cm-\tabcolsep-\widthof{$\Bigg]$}}{air}]& \\
         29                 & $^{15}$N$^{14}$N                      &  \num{>3.1e+01} &No && \\
         30                 & $^{15}$N$_2$, NO                      &  \num{>3.9e+00} &No && \\
         31                 & $^{15}$NO                             &  \num{>6.3e-01} &No & & \\
         32                 & O$_2$                                 &  \num{>1.5e+01} &No && \\
         36                 & $^{36}$Ar                             &  \num{>1.9e+00} &No && \\
         40                 & $^{40}$Ar                             &  \num{>1.1e+02} &No && \\  \cdashline{1-4} \noalign{\vskip 0.5ex}
         44                 & CO$_2$                                &  \num{2.4e+00} &Yes & RGA intrinsic &\\
         \bottomrule
    \end{tabularx}
    \label{tab:rga_spectra_alt}
\end{table}

\subsection{Column Density}

The column density of tritium molecules inside the WGTS source tube is an important parameter for the KATRIN experiment as it directly affects the number of emitted $\upbeta$-electrons, as well as their scattering on the gas. 
At a constant temperature of the WGTS, the main parameter which influences the column density is the flow rate of tritium through the source tube.
This flow rate and the forepressure of the WGTS TMPs are the limiting factors for the column densities which can be achieved.
In the \oldkryptonmode{} the forepressure serves as the injection pressure in the WGTS and limits the column density to about \SI{2e21}{\per\meter\squared} at \SI{80}{\kelvin}. 
In contrast, in the \newkryptonmode{}, the flow through the TMPs limits the column density to a higher level of about \SI{3.75e21}{\per\meter\squared} at the same temperature.

\clearpage

\section{Discussion}
\label{sec:discussion}

\subsection{Spectroscopic Measurements}
\label{sec:spectroscopicmeasurements}
The precise measurement of the parameters which define the spatial inhomogeneities of the plasma potential in the KATRIN tritium source is the motivation for the production of the ultra-strong \Rb{} source that can provide sufficiently large activities of \KrM{} for the \newkryptonmode{} \cite{Machatschek2021}.

The $\upbeta$-decay and subsequent ionization processes generate an abundance of electrons and positive ions in the tritium source, which form a plasma. It is defined by the gas flow, recombination and surface potentials.

The absolute value of the starting potential of $\upbeta$-electrons within this plasma will be observed as an energy offset of the tritium spectrum with regard to the kinematic endpoint. 
A radially different starting potential can be accounted for by allowing different endpoint fit parameters for individual rings in the segmented focal plane detector.
If not considered in the analysis, this would pose a bias for the neutrino mass observable.

Longitudinal inhomogeneities in the source electric potential will manifest themselves as deformation of the recorded tritium spectrum. If not considered in the analysis, to leading order this results in a bias on the neutrino mass of $-2\sigma_P^2$ \cite{Robertson1988}, where $\sigma_P$ is the longitudinal standard deviation of the electric potential.

$\upbeta$-electrons starting in the downstream part of the WGTS mainly leave the source without inelastic scattering off tritium molecules. 
Those starting in the upstream part have to pass through a more dense gas column before reaching the spectrometer section.
Therefore, they undergo scattering off molecules with a characteristic energy loss \cite{eloss2021} at a higher probability. 
The total spectrum is therefore a superposition of the individual spectra for each scattering multiplicity. 
A longitudinal asymmetry in the electric starting potential will therefore not only deform the individual spectra, but will also shift the energy-offsets of  $n$-fold-scattered spectra relative to each other due to different effective starting potentials. 
For KATRIN, most importantly the shift between unscattered and 1-fold scattered electrons needs to be accounted for, which is denoted as $\Delta_P$ \cite{Machatschek2021}.

\begin{figure}
    \centering
    \includegraphics[width=\textwidth]{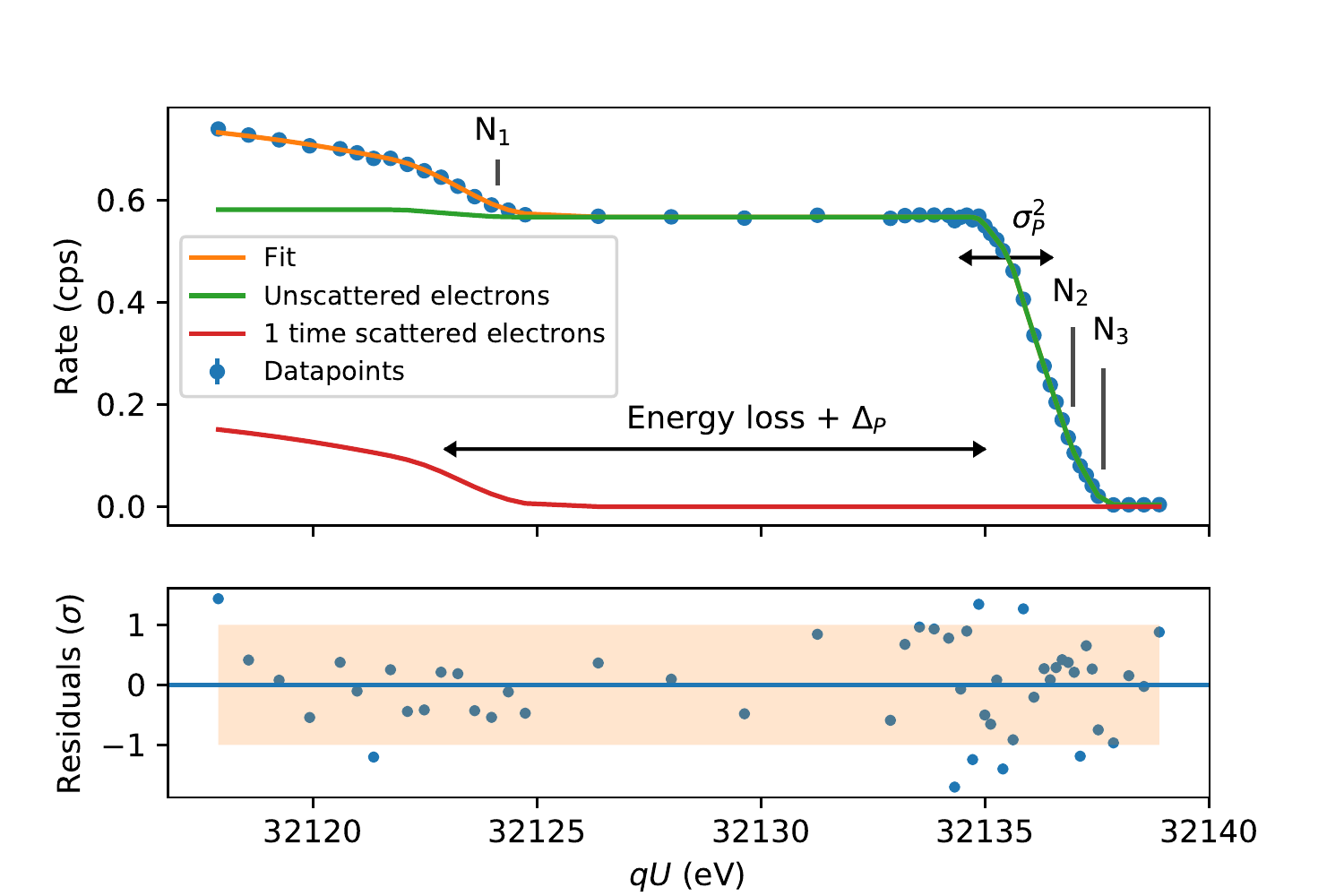}
    \caption{Exemplary spectrum of the $N_1$-32, $N_2$-32, and $N_3$-32 of co-circulating \KrM{} in the WGTS at a column density of $\SI{3.75e21}{\per\metre\squared}$ in the \newkryptonmode{}. 
    The acquisition was \SI{515}{\hour} corresponding to a summation of 180 individual scans of each \SI{\approx 3}{\hour}. 
    For the sake of clarity, only the data of 1 pixel of the FPD is shown.}
    \label{fig:KrNlinespectrum}
\end{figure}

High-resolution electron spectroscopy of the spectral range of the $N_1$-32, $N_2$-32, and $N_3$-32 lines around $\SI{32}{\kilo\electronvolt}$ has been identified as most suitable for the measurement of these parameters due to minimal systematic effects despite the low relative intensity. 
\autoref{fig:KrNlinespectrum} shows an example spectrum for measurements performed during \SI{515}{\hour}. 

The natural line widths of the $N_2$-32 and $N_3$-32 lines are assumed to be vanishing and are not further collision-broadened due to the low gas density in the WGTS. Therefore the line shape of the integral spectrum is only determined by the transmission function of the MAC-filter - thus by the magnetic fields in the source and the analysis plane and by the Gaussian broadening parameter $\sigma_P^2$. The $\Delta_P$ parameter corresponds to an additional energy shift between un-scattered and 1-fold scattered electron spectra. The precision fitting of these spectra are beyond the scope of this paper and will be discussed in an upcoming publication.

The systematic uncertainty on the neutrino mass measurement from the plasma calibration can be determined by propagation of the input parameter uncertainties in the spectral model by Monte-Carlo methods \cite{PhysRevD.104.012005}. 
For illustration purposes it can be approximated for the individual contributions by $\sigma_P^2$ and $\Delta_P$ as 
\begin{align}
    \delta(m_\nu^2)_{\mathrm{broadening}} &= 4\sigma_P\cdot\delta(\sigma_P)~,\\
    \delta(m_\nu^2)_{\mathrm{shift}} &= \epsilon\cdot\delta(\Delta_P)~.
\end{align}
Here $\delta(\sigma_P)$ and $\delta(\Delta_P)$ are the uncertainties of the parameters; $\epsilon$ is an effective parameter which describes the susceptibility of the $\Delta_P$ parameter on the neutrino mass \cite{Machatschek2021}. 
For the current KATRIN measurement configuration it is about $\epsilon \approx\SI{ 1.15}{\electronvolt}$. 
This large value of $\epsilon$ leads to  $\delta(m_\nu^2)_{\mathrm{shift}}$ dominating over $\delta(m_\nu^2)_{\mathrm{broadening}}$ by far.

\begin{figure}
    \centering
    \includegraphics[width=\textwidth]{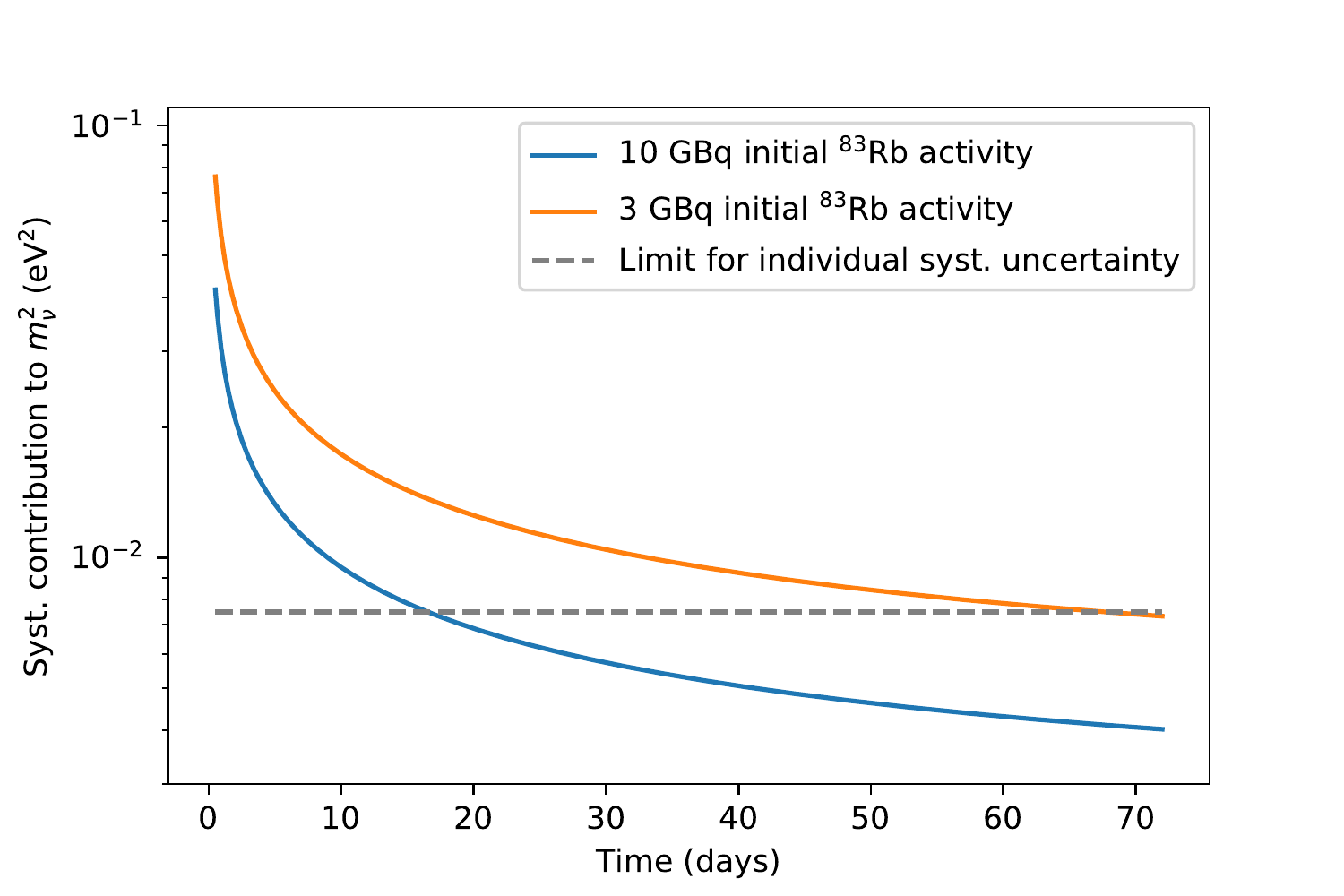}
    \caption{Systematic contribution from the plasma observable uncertainty $\delta(\Delta_P)$ on the neutrino mass parameter $m_{\nu}^2$ as a function of the measurements time, calculated for two possible \Rb{} source activities. The sensitivity is given for a uniform value for the entire detector.}
    \label{fig:Sensitivity}
\end{figure}

The uncertainty in the determination of the plasma observables strongly depends on the available measurement time, which is depicted in figure \autoref{fig:Sensitivity}. 
It shows the evolution of $\delta(\Delta_P)$ with time as calculated for the actual krypton-line signal strength as recorded in the \newkryptonmode{} with the strong \SI{10}{\giga\becquerel} \Rb{} source.

In order to meet the requirements of KATRIN for reaching the target neutrino mass sensitivity, it is relevant that all the systematic effects add up to less than $\SI{1.7e-2}{\electronvolt\squared}$, which implies that individual systematic effects like the starting potential inhomogeneity should be less than $\SI{7.5e-3}{\electronvolt\squared}$.

Expected systematic contributions to the neutrino mass are plotted into \autoref{fig:Sensitivity} for two different \Rb{} source activities. It shows that one would need to measure more than two months to reach the required precision on the dominating $\Delta_P$ parameter with a medium strong source of $\SI{3}{\giga\becquerel}$. 
It is demonstrated that only a \Rb{} source with an activity on the order of \SI{10}{\giga\becquerel}, or more, can be used to determine plasma observables with the required accuracy to fulfill the requirements of KATRIN systematics budget within reasonable time scales.

\subsection{Observed \Rb{} Half-Life}
\label{sec:kr_halflife}

\begin{figure}
    \centering
    \includegraphics[width=\textwidth]{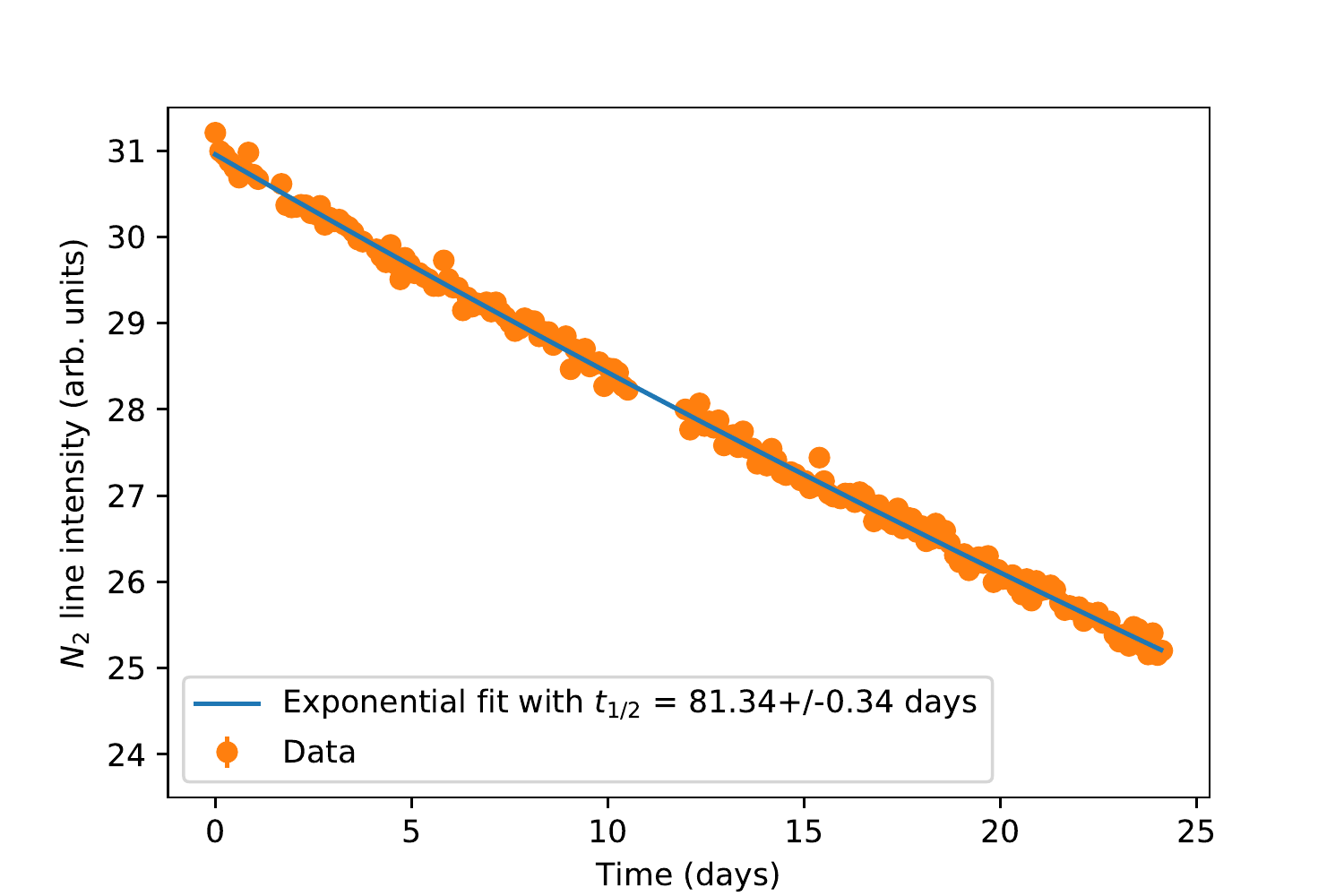}
    \caption{Evolution of the intensity of the $N_2$-32 \KrM{} line as a function of time.}
    \label{fig:Activity-of-N23}
\end{figure}

The trend of the $N_2$-32 line intensity as a function of time is shown in \autoref{fig:Activity-of-N23}. It is decreasing with a half-life of \SI{81.34\pm0.34}{\day} which is significantly shorter than the value of \SI{86.2\pm0.1}{\day} reported in literature \cite{Audi2003} for the \Rb{} half-life.
Based on this deviation, we conclude that the \KrM{} in the system, and specifically inside the WGTS tube (the source of the measured $N_2$-32 conversion electrons), is not in the expected equilibrium with its \Rb{} parent. 
Possible reasons for this phenomenon are a decrease of the \KrM{} emanation rate from the zeolite source over time or a time dependent increase in loss of \KrM{} in the loop system. 
Nevertheless, the observed decrease is slow enough to enable the measurement of the scans of the conversion electrons, even without the count correction on this decrease during the scan.
As the main intention for using \KrM{} in KATRIN is as a calibration source, which is possible with this observed half-life discrepancy, it is currently not planned to spend extensive amounts of measurement time in order to investigate this effect in detail.

\section{Summary}
The KATRIN experiment requires detailed knowledge of the starting conditions of $\upbeta$-electrons produced in tritium decay in its Windowless Gaseous Tritium Source.
In order to investigate these conditions, mono-energetic conversion electrons from the decay of gaseous \KrM{} can be used by mixing it with the tritium.
This publication describes the two operation modes of the KATRIN experiment with gaseous \KrM{} as well as the achieved rates and their stability.
Based on these results, it is shown that a sensitivity on the starting conditions needed to meet the KATRIN requirements for the target sensitivity on the neutrino mass of \SI{0.2}{\eV}, can be reached in a two-week long measurement with a \SI{10}{\giga\becquerel} \Rb{} activity source in the \KrM{} generator. 
Such a source (\SI{9.8}{\giga\becquerel}) was used during the June - August 2021 krypton measurement campaign of the KATRIN experiment. 
The results of this measurement campaign will be presented in upcoming publications.

\acknowledgments
We acknowledge the support of Helmholtz Association (HGF), Ministry for Education and Research BMBF (05A17PM3, 05A17PX3, 05A17VK2, 05A17PDA, and 05A17WO3), Helmholtz Alliance for Astroparticle Physics (HAP), the doctoral school KSETA at KIT, and Helmholtz Young Investigator Group (VH-NG-1055), Max Planck Research Group (MaxPlanck@TUM), and Deutsche Forschungsgemeinschaft DFG (Research Training Groups Grants No., GRK 1694 and GRK 2149, Graduate School Grant No. GSC 1085-KSETA, and SFB-1258 in Germany; 
Ministry of Education, Youth and Sport (CANAM-LM2015056, LTT19005) in the Czech Republic; 
and the Department of Energy through grants DE-FG02-97ER41020, DE-FG02-94ER40818, DE-SC0004036, DE-FG02-97ER41033, DE-FG02-97ER41041,  {DE-SC0011091 and DE-SC0019304 and the Federal Prime Agreement DE-AC02-05CH11231} in the United States. 
This project has received funding from the European Research Council (ERC) under the European Union Horizon 2020 research and innovation programme (grant agreement No. 852845). 
We thank the computing cluster support at the Institute for Astroparticle Physics at Karlsruhe Institute of Technology, Max Planck Computing and Data Facility (MPCDF), and National Energy Research Scientific Computing Center (NERSC) at Lawrence Berkeley National Laboratory.

We thank Radom\'{i}r B\v{e}hal, Petr Han\v{c}, Daniel Seifert, V\'{a}clav Zach, and Jan \v{S}tursa for their help in preparing the Rb source.

\paragraph{Author contributions}

Operation of the KATRIN experiment and data taking during the \KrM{} measurement campaigns evaluated in this paper — all members of the KATRIN Collaboration.
Data evaluation: Alexander Marsteller, Magnus Schl\"{o}sser, Matthias B\"{o}ttcher.
Writing--original draft preparation: Alexander Marsteller, Beate Bornschein, Ond\v{r}ej Lebeda, Florian Priester, Jan R\'{a}li\v{s}, Carsten R\"{o}ttele, Magnus Schl\"{o}sser, Michael Sturm, Drahoslav V{\'{e}}nos.
Writing--review and editing: 
Reviewed by all members of the KATRIN Collaboration, edited by Alexander Marsteller and Magnus Schl\"{o}sser.
All authors have read and agreed to the published version of the manuscript.

\bibliographystyle{JHEP}
\bibliography{literature.bib}

\end{document}